\documentclass[aps,prl,twocolumn,groupedaddress,showpacs]{revtex4}
\usepackage[latin1]{inputenc}
\usepackage[english]{babel}
\usepackage{setspace}  
\usepackage{graphicx}
\usepackage{amssymb}
\usepackage{amscd}
\usepackage{rotating}
\usepackage{color}
\begin{document}

\preprint{}

\title{Vacancy supersolid of hard-core bosons on the square lattice}

\author{Long  Dang$^1$, Massimo Boninsegni$^1$ and Lode Pollet$^2$}
\affiliation{$^1$Department of Physics, University of Alberta, Edmonton, Alberta, Canada, T6G 2G7
\\
$^2$Theoretische Physik, ETH Z\"urich, 8093 Z$\ddot{u}$rich, Switzerland}

\date{\today}

\begin{abstract}
The ground state of hard-core bosons on the square lattice with nearest and next-nearest neighbor repulsion is studied by Quantum Monte Carlo simulations. A supersolid phase with vacancy condensation and 'star' diagonal ordering is found for filling $\rho < 0.25$. At fillings $\rho$ $>$ 0.25 a supersolid phase exists between the star and the stripe crystal at $\rho$=0.5. No supersolid phase occurs for $\rho$ $>$ 0.25 if the ground state at half-filling is either a checkerboard crystal or a superfluid.  No commensurate supersolid phase is observed. 
\end{abstract}

\maketitle

Experimental advances in the field of ultracold atoms in optical lattices \cite{jaksch9805} have given renewed impetus to the investigation of novel phases of matters, especially those displaying simultaneously different types of order. One such phase is the {\it supersolid}, featuring both diagonal and off-diagonal long range order, which has been the subject of much theoretical speculation \cite{group1}. Recent claims of observation of this phase in solid helium \cite{kim0405} have generated some controversy \cite{prokofev07}.

Theoretical studies have yielded strong evidence of supersolid phases of lattice bosons, for various types of  model interactions among atoms, as well as of lattice geometries \cite{square,sengupta,melko,group03}.
In the presence of nearest-neighbor particle hopping and on-site hard core repulsion, the supersolid phase  is only observed on the {\it interstitial} side of a commensurate solid phase (e.g., for 1/3 $<$ $\rho$ $<$ 2/3 on the triangular lattice) \cite{group03}. In contrast, doping with vacancies results in the coexistence of an insulating  crystal and a superfluid by the formation of a domain wall~\cite{note2}. Phase separation of vacancies is also observed in {\it ab initio} simulations of helium crystals \cite{kuklov}. This seems remarkable, as Bose condensation of vacancies has long been regarded as the paradigm for supersolidity \cite{group1}.  The purpose of this work is to gain understanding in the asymmetry between the behavior of vacancies and interstitials, and explore physical conditions that underlie a vacancy supersolid phase in lattice bosons.

Our starting point is the well-known  Hamiltonian
\begin{eqnarray}\label{one}
\hat H= &-&t
\sum_{\langle ij\rangle}
(\hat{b}_i^{\dagger}\hat{b}_j + h.c.)  + V_1\sum_{\langle ij\rangle}\hat{n}_i\hat{n}_j \nonumber \\ &+& 
V_2\sum_{\langle\langle ik\rangle\rangle}\hat{n}_i\hat{n}_k - \mu \sum_i \hat{n}_i .
\end{eqnarray}
A square lattice with periodic boundary conditions of $N=L\times L$ sites is assumed. The boson density (filling) is $\rho=N_B/N$, where the number of particles $N_B$ is determined by the chemical potential $\mu$. 
The operator $\hat{b}^\dagger_i(\hat{b}_i)$ creates (annihilates) a hard-core boson on site $\it{i}$, with a maximum occupation number $\hat n_i=\hat b_i^\dagger \hat b_i$  of one particle per site.
The first term of (\ref{one}) describes particle hopping to a nearest-neighboring site with amplitude $t$, which is our energy scale, $t=1$. The second and third terms represent repulsive interactions between bosons on nearest  and next-nearest neighboring sites, respectively. 

\begin{figure}[h]
\includegraphics[scale=0.38,angle=0]{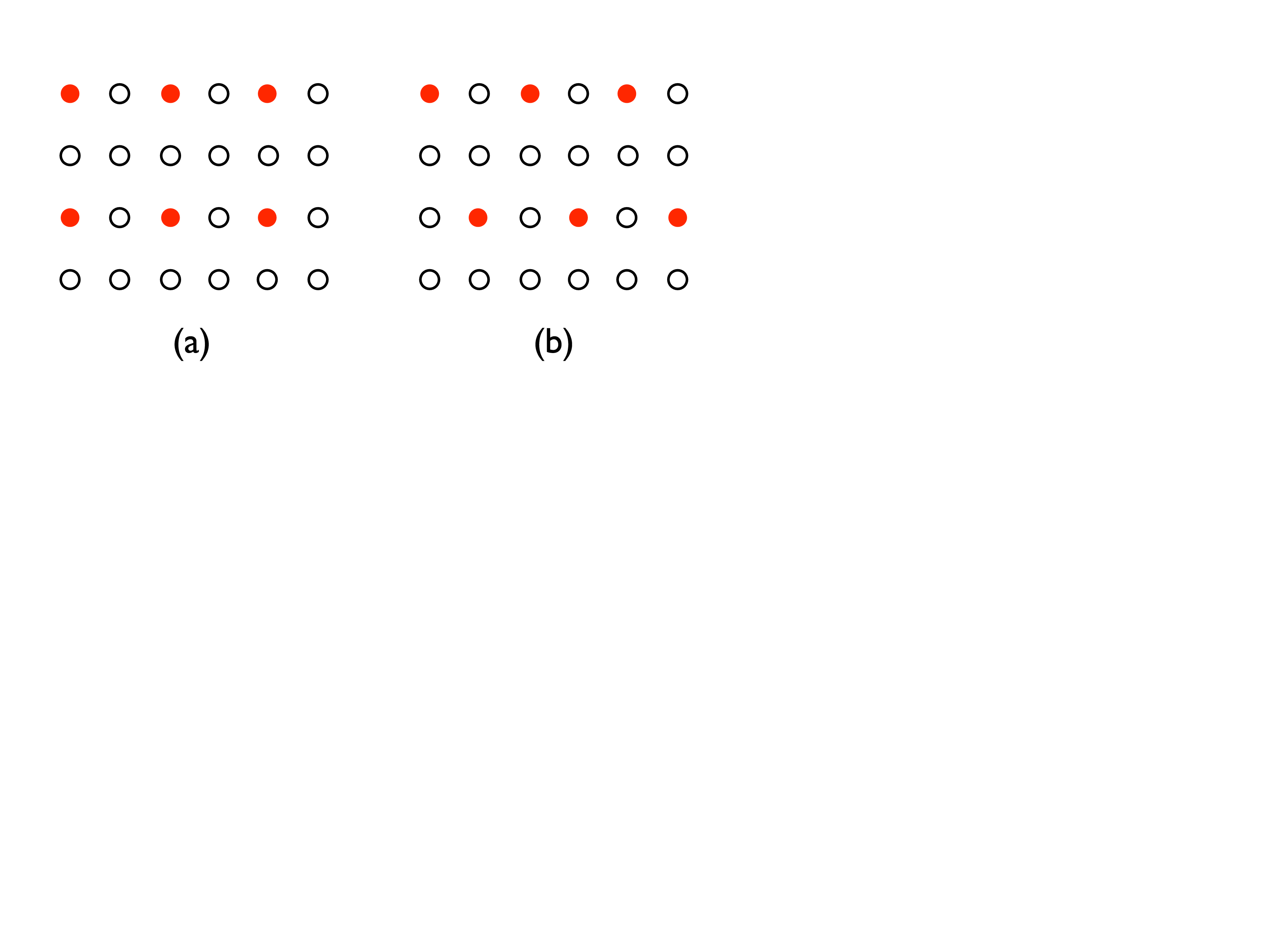}
\caption{ (Color online). Classical, degenerate star crystal phases at quarter-filling on the square lattice. Filled circles represent particles, empty circles empty sites. Quantum fluctuations lift the degeneracy, selecting ({\it a}) as the ground state if $V_1 < 2 V_2$, ({\it b}) otherwise. }\label{refm1}
\end{figure}

Previous studies of the ground state of (\ref{one}) have yielded evidence of  three possible phases at half-filling ( $\rho$=0.5):  a superfluid, a checkerboard solid and a stripe solid. The latter two are commensurate and insulating phases. Doping the stripe crystal away from half filling yields a supersolid phase, whereas a first-order quantum phase transition separates the checkerboard crystal from a superfluid  \cite{square}. 
For sufficiently strong next-nearest neighbor repulsion, the Hamiltonian (\ref{one}) also features a commensurate crystal phase at 
quarter \cite{note} filling, referred to as ``star" (Fig. \ref{refm1}).  It has also been shown that  a supersolid phase can occur, upon doping the star solid above quarter-filling \cite{melko}. 

In this work, we systematically investigated the ground state phase diagram of (\ref{one}).
Our main findings are: \\
{\it i}) Doping a star solid with vacancies {\it always} gives rise to a supersolid phase.\\
{\it ii}) Between a star and a stripe solid a supersolid phase can be observed; phase transitions are  continuous. \\
{\it iii}) If the phase at half filling is not a striped solid but a checkerboard solid or a superfluid, no supersolid phase exists.  First order phase transitions separate superfluid from crystal phase(s).\\
{\it iv}) In agreement with Ref. \onlinecite{melko}, we find no {\it commensurate} supersolid phase \cite{notep}. 

Our Monte Carlo simulations are based on the Worm Algorithm in the lattice path-integral representation \cite{prokofev98}, using the implementation described in Ref. \onlinecite{lode07}. The simulations are at low temperature (typically $\beta=L$), in order to extract the properties of the system in its ground state. 
\begin{figure}[h]
\includegraphics[scale=0.30,angle=0]{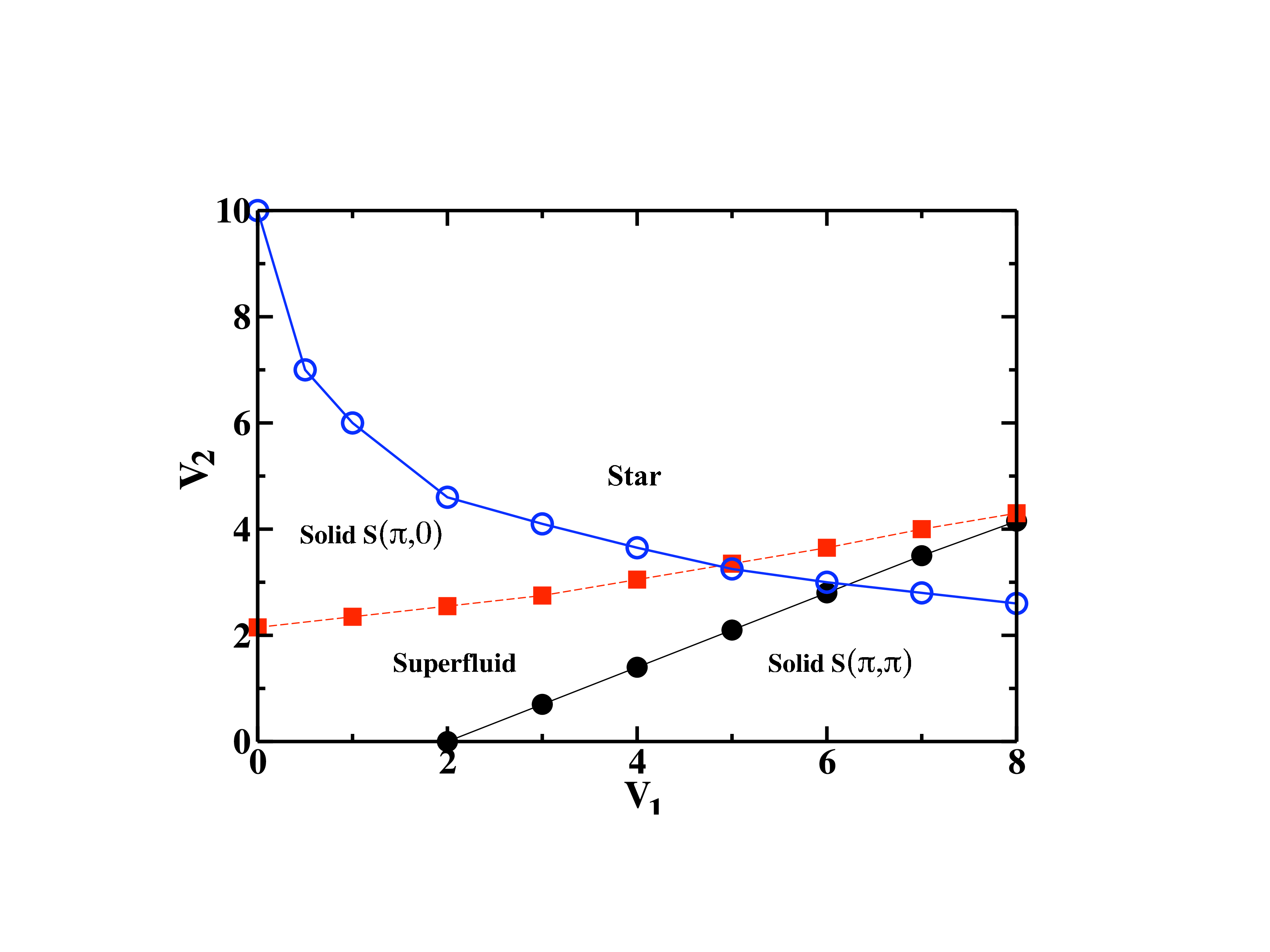}
\caption{ (Color online). Ground state phase diagram of (\ref{one}). Lines are guides to the eye. Symbols lie at  computed phase boundaries, namely between {\it a}) a ``star" crystal and a superfluid at $\rho=0.25$ (open circles) {\it b})  a ``striped" $(\pi,0),(0,\pi)$ crystal and a superfluid at $\rho=0.5$ (filled boxes) and {\it c}) a superfluid a ``checkerboard" crystal (filled circles), also at $\rho=0.5$. Statistical errors are smaller than symbol sizes. Results at $\rho=0.5$ are consistent with previous works \cite{square}.}\label{ref0}
\end{figure}

In order to characterize the various phases, we computed the superfluid fraction $\rho_S$ using the well-known winding number estimator \cite{ceperley848687},  as well as the static structure factor 

\begin{eqnarray}
S({\bf Q})= \frac{1}{N^2}\biggl \langle\biggl | \sum_{i=1}^{N}\hat{n}_i\ e^{i{\bf Q}\cdot{\bf r}_i}  \biggr |^2\biggr \rangle , 
\end{eqnarray}
where $\langle\dots\rangle$ stands for thermal average. Presence of crystalline long-range order is signaled by a finite value of $S({\bf Q})$ for some specific wave vector. In particular, ${\bf Q} = (\pi,\pi)$ is the wave vector associated to checkerboard order at half filling, whereas  ${\bf Q} = (\pi,0)$, $(0,\pi)$ signals striped order at half filling, as well as star order at quarter filling.

\begin{figure}
\includegraphics[scale=0.30,angle=0]{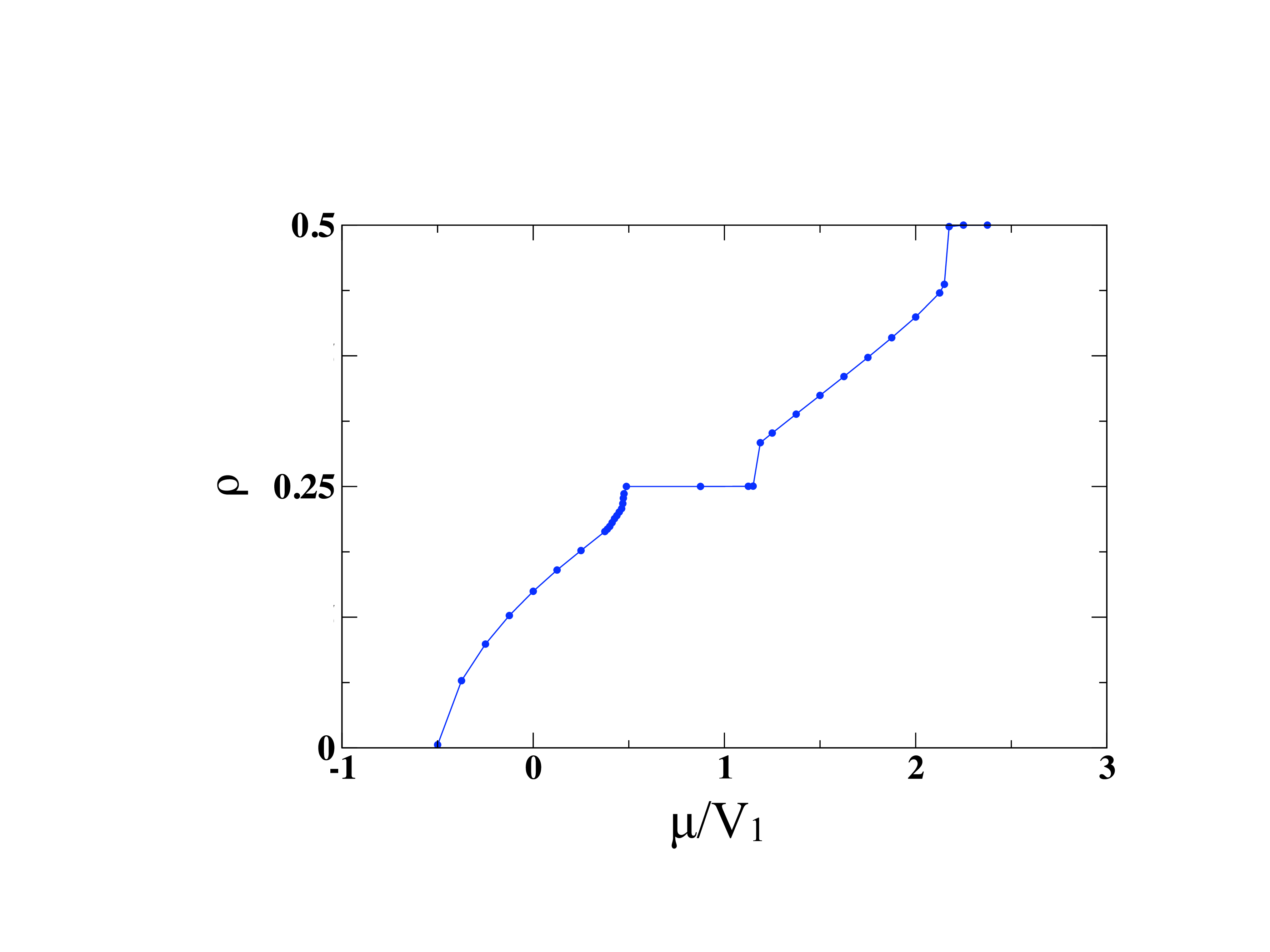}
\caption{(Color online). Density $\rho$ versus chemical potential $\mu$ for $V_1=8$, $V_2=3.5$. Statistical errors are smaller than symbol sizes. A first-order quantum phase transition separates the ``star" crystal at $\rho=0.25$ and the ``checkerboard" crystal at $\rho=0.5$ from a superfluid phase at intermediate densities. Results shown are for a square lattice with $L$=24. A continuous phase transition to a vacancy supersolid occurs below $\rho=0.25$.} \label{f2a}
\end{figure}

\begin{figure}
\includegraphics[scale=0.30,angle=0]{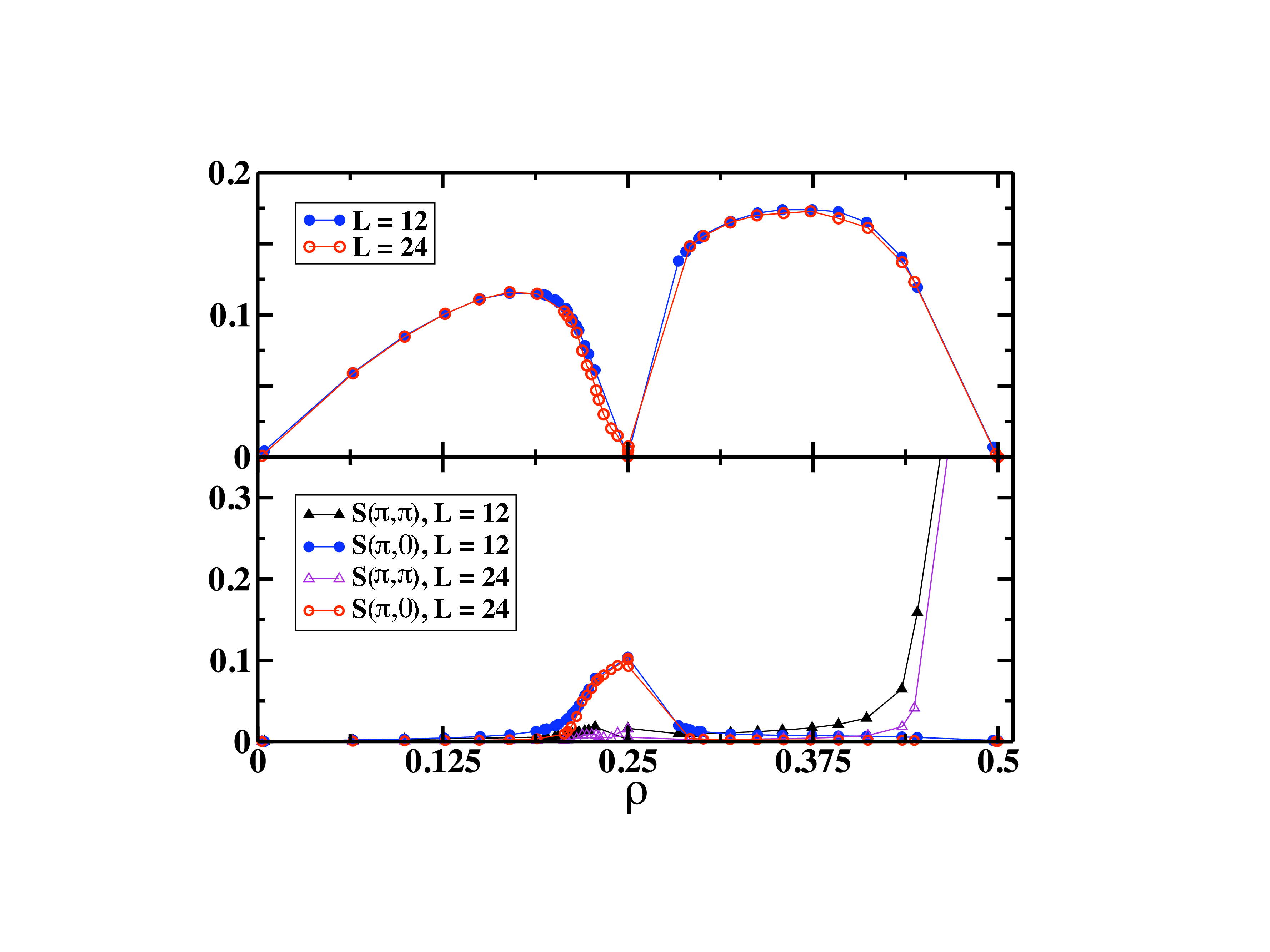}
\caption{(Color online). Superfluid density $\rho_S$ (upper panel) and static structure factor $S({\bf Q})$ (lower panel) for $V_1=8$, $V_2=3.5$ for two system sizes and the two wave vectors ${\bf Q}=(\pi,\pi)$ and ${\bf Q}=(\pi,0),(0,\pi)$. Statistical errors are smaller than symbol sizes.} \label{f2b}
\end{figure}

Fig. \ref{ref0} summarizes our results for the ground state phase diagram of (\ref {one}). Three distinct phase boundaries are identified. The first one (open circles) separates a superfluid from  a star crystal at quarter filling. In this work, we focus our attention on the region {\it above} the open circles, i.e., wherein a star crystal exists. The other two phase boundaries refer to the behavior of the system at half-filling,  separating a stripe crystal (filled boxes) [checkerboard crystal (filled circles)]  from a superfluid. We will now discuss the physical behavior at other densities as a function on $V_1$ and $V_2$.


Fig. \ref{f2a} shows the density computed as a function of the chemical potential for $V_1=8$, $V_2=3.5$. The ground state of the system is a checkerboard crystal at half filling. Density jumps signal first-order phase transitions above quarter- and below half-filling; everywhere else the curve is continuous albeit with a clear discontinuity of the first derivative (i.e., of the compressibility) on approaching $\rho=0.25$ from below.  Fig. \ref{f2b} displays the superfluid density as well as the static structure factor, computed for both ($\pi$,0) and ($\pi$,$\pi$). The superfluid density is everywhere finite, except at commensurate fillings, whereas the static structure factor is finite at and near commensuration. 

\begin{figure}[t]
\includegraphics[scale=0.30,angle=0]{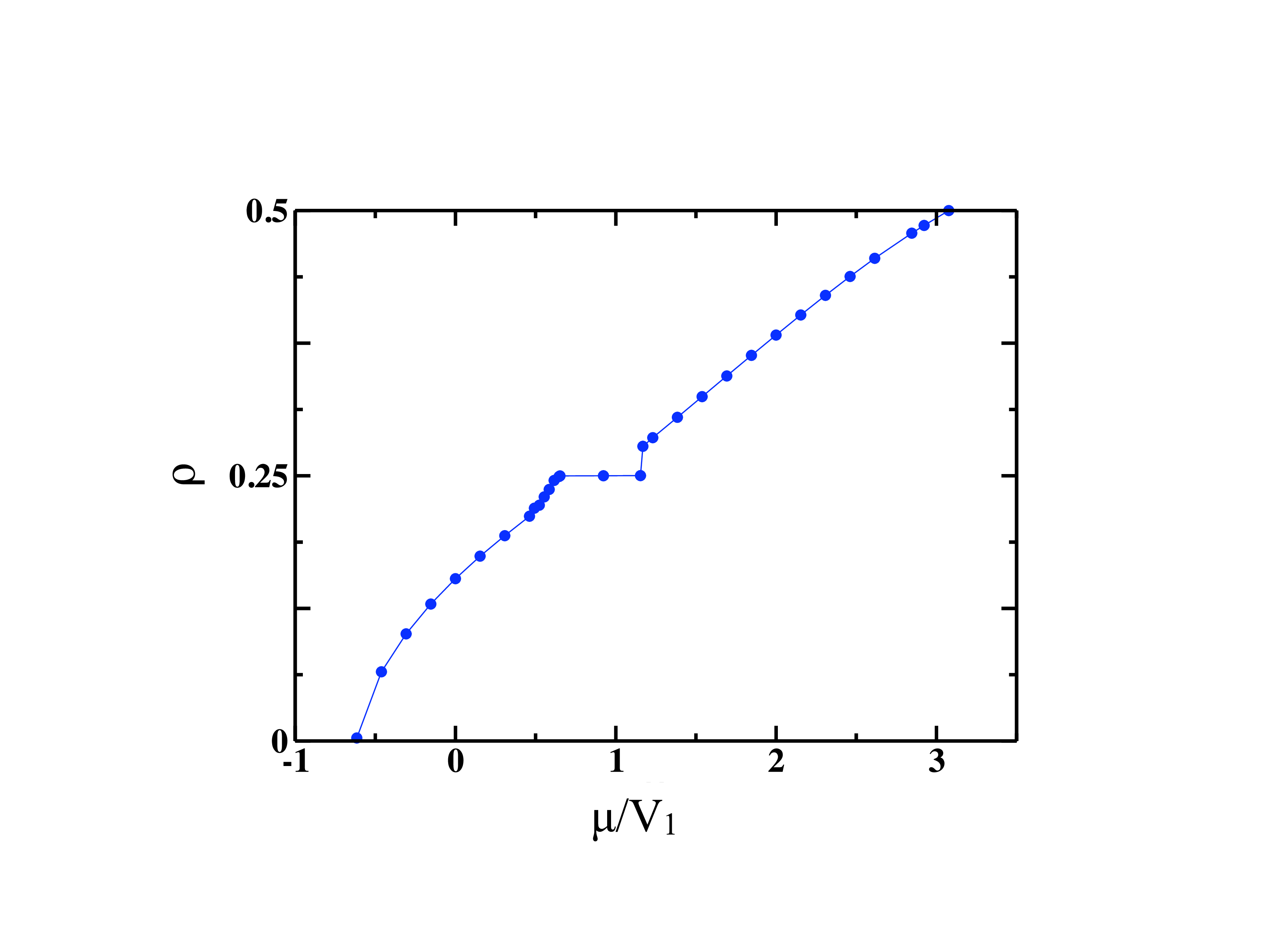}
\caption{(Color online). Density $\rho$ versus chemical potential $\mu$ for $V_1=6.5$, $V_2=3.5$ and linear system size $L=24$. Statistical errors are smaller than symbol sizes. A continuous quantum phase transition separates the star crystal from a supersolid on the vacancy side ( $\rho $ $<$ 0.25), whereas a first-order phase transition from the stripe crystals to a  superfluid exists on the interstitial ($\rho$ $>$ 0.25) side. } \label{f3p}
\end{figure}
\begin{figure}[h]
\includegraphics[scale=0.30,angle=0]{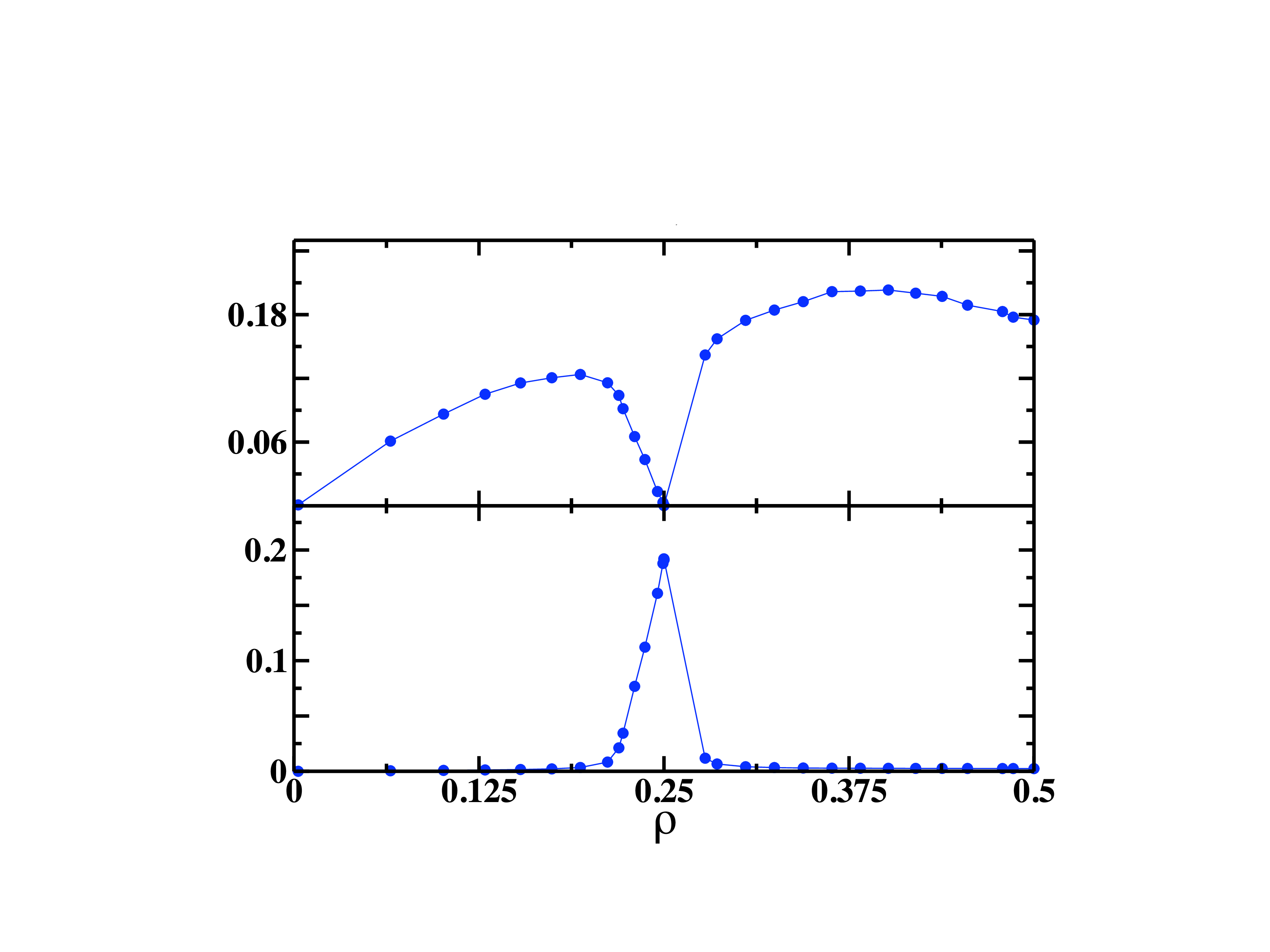}
\caption{(Color online). Superfluid density $\rho_S$ (upper panel) and static structure factor $S({\bf Q}=(\pi,0),(0,\pi))$ (lower panel) for $V_1=6.5$, $V_2=3.5$ for two system sizes.  System size is $L=24$. A supersolid phase exists only on the vacancy side below quarter-filling. The ground state of the system at half-filling is superfluid.} \label{f5p}
\end{figure}

Based on these results, we conclude that the system is superfluid for 0.25 $<$ $\rho$ $<$ 0.5, with first-order quantum phase transitions to the star (checkerboard) crystal at $\rho=0.25$ ($\rho=0.5$). For $0 \le \rho \le 0.25$, on the other hand, the system features two continuous phase transitions,  one from a superfluid to a {\it vacancy-based supersolid}, corresponding to the change in slope of the $\rho(\mu)$ curve, and the other from the supersolid to the star crystal at $\rho=0.25$. It is interesting to compare the results right above and below quarter filling (Fig. \ref{f2b}) in order to appreciate the effectiveness in distinguishing a homogeneous supersolid phase from coexistence of two phases.

A  similar scenario is observed for ($V_1$,$V_2$) pairs for which the ground state of the system at half-filling is superfluid, where the only difference is the absence of the crystalline phase at $\rho=0.5$. In this case too, a vacancy-based supersolid exists below quarter-filling and a first-order phase transition separates the star crystal from a superfluid above quarter-filling. The superfluid phase extends all the way to half-filling as the competition between nearest-neighbor and next-nearest-neighbor interactions does not result in any crystal ordering. Results for one example in this region, i.e., $V_1=6.5$, $V_2$=3.5, are shown in Figs. \ref{f3p} and \ref{f5p}. 

\begin{figure}
\includegraphics[scale=0.30,angle=0]{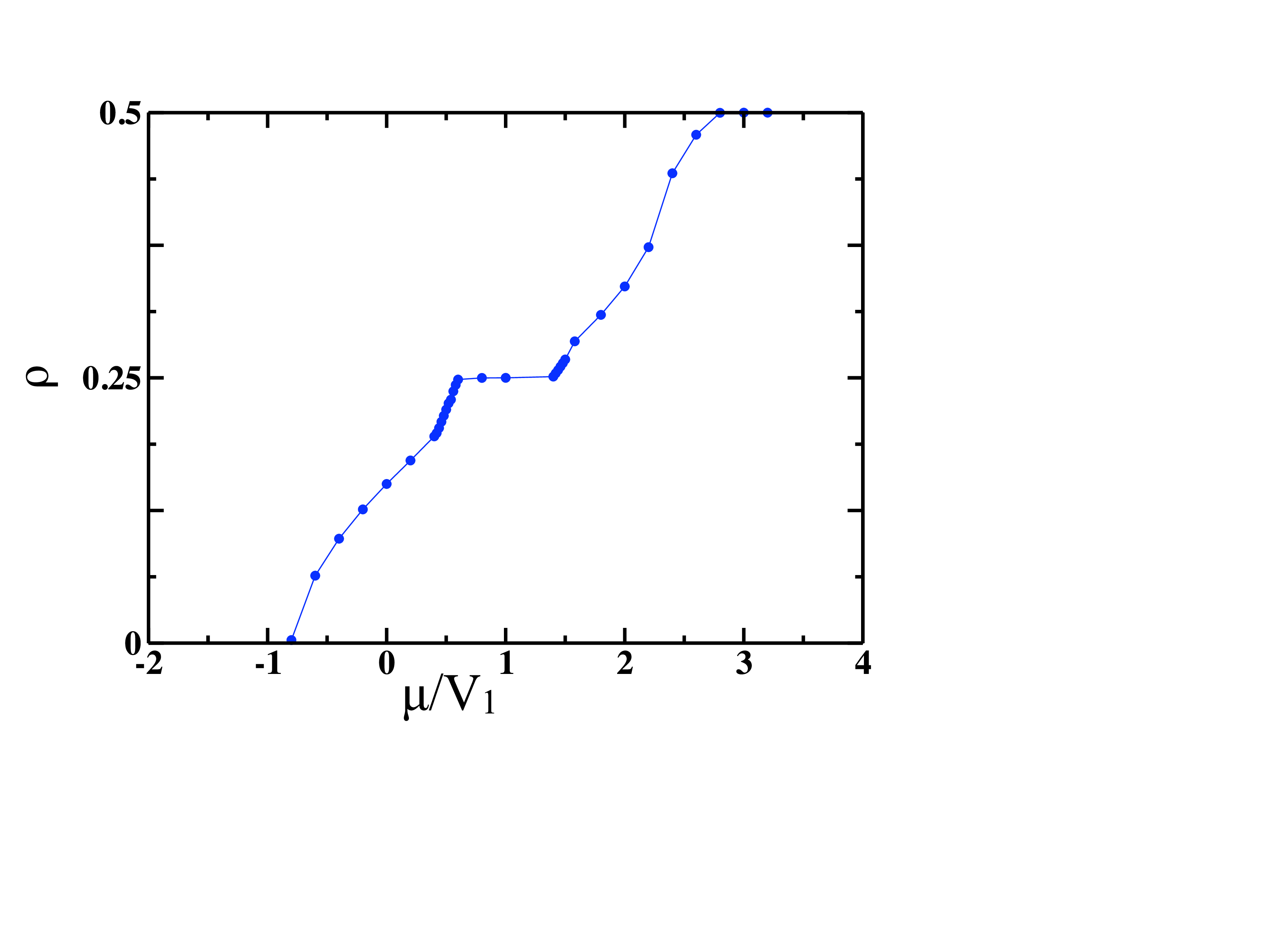}
\caption{(Color online.) Density $\rho$ versus chemical potential $\mu$ for $V_1=5$, $V_2=5$ and linear system size $L=24$. Statistical errors are smaller than symbol sizes. Continuous quantum phase transitions separate both star and stripe crystals from supersolid phases on both sides. A continuous phase transition  from a supersolid to a superfluid takes place below quarter-filling.  } \label{f7}
\end{figure}

\begin{figure}
\includegraphics[scale=0.30,angle=0]{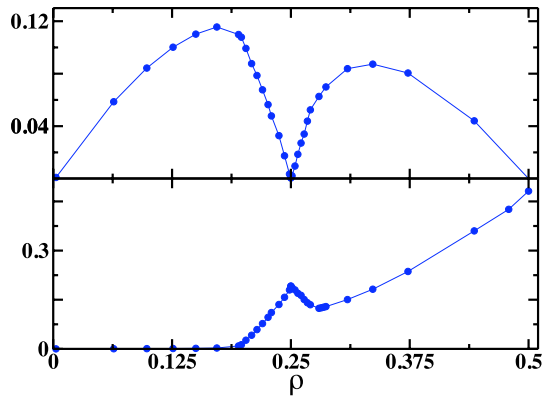}
\caption{(Color online). Superfluid density $\rho_S$ (upper panel) and static structure factor $S({\bf Q}=(\pi,0),(0,\pi))$ (lower panel) for $V_1=5$, $V_2=5$.  System size is $L=24$. A supersolid phase exists everywhere between quarter- and half-filling, and below quarter-filling.} \label{f8}
\end{figure}

This behavior should be contrasted to the one shown in Figs. \ref{f7} and \ref{f8} for the case $V_1=V_2=5$. This parameter choice stabilizes a star crystal at quarter-filling and a stripe crystal at half-filling.  A supersolid phase exists everywhere between the star and the stripe crystal. A vacancy-rich supersolid phase occurs below filling 0.25, with a continuous phase transition to a superfluid at lower densities. A phase transition also occurs in this case above quarter filling, between a star and a striped supersolid; its location is signaled by a ``kink"  in the numerical value of the static structure factor (Fig. \ref{f8}, lower panel). This phase transition was claimed to be of first order in Ref. \onlinecite{melko}, based on a discontinuous jump in the value of the quantity $|S(\pi,0)-S(0,\pi)|$ across the transition. 
If the values of $V_1$ and $V_2$ are such that a checkerboard crystal or a superfluid emerges at half-filling no supersolid phase intervenes in the $0.25 < \rho < 0.5$ interval. 


The physical mechanism underlying the presence of a supersolid phase in this model away from commensurate fillings, even on lattices other than the square, is whether or not defects such as interstitials or vacancies can move without frustration.
On the triangular lattice at filling $\rho$=1/3, and in the presence of nearest-neighbor repulsion only, interstitial particles can freely hop around, hence Bose-condensing. At a finite concentration of vacancies, on the other hand, the system can lower its energy by forming a domain wall, i.e., vacancies phase separate~\cite{group03}. The roles are reversed in our case, on the square lattice at quarter-filling. There is no cost in moving a vacancy around, in either the Fig. \ref{refm1}({\it a}) or \ref{refm1}({\it b}) scenarios.  Vacancies will thus condense. Conversely, the lowest energy needed to create an interstitial in the case of Fig. \ref{refm1}({\it b}) (i.e., $V_1 > 2 V_2$)  is $4 V_2$, which requires that a crystal atom also move to an empty row. The energy can only be lowered by higher order processes in $t/V_1$, which gives rise to the same domain wall argument invoked above for vacancies. This is the phase separation shown in Fig. \ref{f2a}. 

The lowest energy needed to create an interstitial in the case shown in Fig. \ref{refm1}({\it a}) (i.e., for $V_1 < 2 V_2$ )  is $2V_1$, which corresponds to an interstitial between two crystal atoms. Those two crystal atoms can now hop around at no potential energy cost, which corresponds to the supersolidity seen in Fig. \ref{f7}.
This simple perturbative argument fails to account for the scenario of  Fig. \ref{f3p}, as higher order processes renormalize the transition point.
The same physical considerations also easily explain the lack of a supersolid phase of either vacancies or interstitials on the {\it kagom\'e} or honeycomb lattices \cite{many}.
 
In an optical lattice, interactions among nearest and next-nearest neighboring atoms could arise from long-ranged (e.g., dipolar) interactions among particles.
The ratio between $V_1$ and $V_2$ for polar molecules is $2\sqrt(2)$. The star solid, and the corresponding vacancy supersolid might thus be seen in experiment for $V_1 > 10$ (see Fig. 2) ~\cite{Innsbruck06}, albeit with rather strong requirements on the density and temperature. Observation of a supersolid seems thus  possible with existing techniques. 
\\ \indent
This work was supported in part by  the Natural Science
and Engineering Research Council of Canada under research grant G121210893, by the Alberta Informatics Circle of Research Excellence and by the Swiss National Science Foundation. Part of the simulations ran on the Hreidar cluster at ETH Zurich.


\begin{thebibliography}{99}
\bibitem{note2} The distinction of vacancies and interstitials is only meaningful away from half-filling.
\bibitem{note} The Hamiltonian (\ref{one}) features particle-hole symmetry. Thus, phase diagram between half and full filling is the specular reflection of that between zero and half filling.
\end{thebibliography}
\end{document}